\documentclass{aastex}

\usepackage{spr-astr-addons}

\begin{document}

\title{The Point of Origin of the Radio Radiation from the Unresolved Cores of Radio-Loud Quasars}

\author{M.B. Bell\altaffilmark{1} and S.P. Comeau\altaffilmark{1}} 

\altaffiltext{1}{Herzberg Institute of Astrophysics,
National Research Council of Canada, 100 Sussex Drive, Ottawa,
ON, Canada K1A 0R6; morley.bell@nrc-cnrc.gc.ca}

\begin{abstract}

Locating the exact point of origin of the core radiation in active galactic nuclei (AGN) would represent important progress in our understanding of physical processes in the central engine of these objects. However, due to our inability to resolve the region containing both the central compact object and the jet base, this has so far been difficult. Here, using an analysis in which the lack of resolution does not play a significant role, we demonstrate that it may be impossible even in most radio loud sources for more than a small percentage of the core radiation at radio wavelengths to come from the jet base. We find for 3C279 that $\sim85$ percent of the core flux at 15 GHz must come from a separate, reasonably stable, region that is not part of the jet base, and that then likely radiates at least quasi-isotropically and is centered on the black hole. The long-term stability of this component also suggests that it may originate in a region that extends over many Schwarzschild radii.
 

\end{abstract}


\section{Introduction}

Jets in AGNs, which include radio galaxies, BL Lacs and quasars, are assumed to be launched from a region close to the central black hole through some process involving power extraction from the black hole spin \citep{bla77,mac82}, or the accretion disk \citep{bla82}. However, few details of this process are known and today, after more than thirty years, the jet formation process in AGNs still remains one of the unsolved fundamental problems in astrophysics \citep{mei01}. One of the main distinguishing characteristics of radio-loud AGNs is their strong core component. However, unlike the material in the jet that can be seen to move outward at high speeds, the position of the core does not move with time; at least no more than can be explained by the fact that the base of the active jet also lies inside the telescope beam. This suggests that the core is more likely to be associated with the central compact object than with the jet. Furthermore, although all core-dominant sources contain a core, they may not all have active jets, and the core cannot be part of the jet if the source has no jet. \citet{blu07} have argued for a thermal origin for the cores in radio quiet quasars and have demonstrated that optically thin bremsstrahlung from a slow, dense disk wind can make a significant contribution to the radio core emission.  Also, no evidence for a jet was found by \citet{kel98,kel04} in several radio loud quasars they mapped (0007+106, 0235+164, 0642+449, 1328+254, 1638+397, 2145+067), and previous work has also suggested that perhaps no more than $\sim$40-70 percent of radio quasars currently have active jets \citep[and references therein]{jia09}. Although all AGNs are likely to experience jet activity at some time, if there are periods when no jet activity is present it would be difficult to explain the core during these times if it were part of the jet. The lack of motion of the core, and the possibility that some sources may not have active jets, are both arguments suggesting that most of the core radiation may be associated with the central compact object and not the jet. However, most investigators still assume that the core radiation from radio loud quasars comes from the innermost portion of the jet, or jet base (JB) \citep[and references therein]{chi99,mar08,fal09}.

A similar investigation discussing the cores of blazars was recently carried out by \citet{mar08} where it was assumed that the core is part of the jet and it was concluded that it was located either where the optical depth in the jet is of order unity, or at a standing shock located somewhere further downstream. It is not clear just how this standing shock would be held in place when everything else in the jet is moving out at high speeds. Perhaps this brightening occurs at a location at the base of the jet where the magnetic field is particularly high; however, this would have to be a general feature of these sources since all have strong cores. It is thus unlikely to be related to the unique HST-1 object seen in the jet of M87 \citep{che07}.

From the multi-frequency flux monitoring results on 3C279 \citep{jor09} there is evidence that opacity is also involved and that we are seeing much deeper into the central engine at short wavelengths. There is evidence that when a new ejection event occurs it is seen first at the shortest wavelengths, showing up later, and presumably further out in the jet, at longer wavelenths. In BL Lac, time delays of $\sim100$ days have been detected between the optical and hard radio events \citep{vil09}. However, as in most other related investigations, it is difficult to ascertain exactly what is happening when the nuclear region and the JB both fall inside the telescope beam.

Polarization has also been used to investigate the nature of the core radiation from these objects, but these observations also must be highly affected by the lack of resolution. The jet radiation is thought to be synchrotron \citep{kro99} and is known to be highly polarized. This can be explained if the jet magnetic field structure is highly organized. This is not the case for the core radiation in most sources where there is little evidence for polarization, which may imply that the region producing the core radiation is highly turbulent. However, this difference can also be considered as further evidence that the core and jet radiation originate from two separate locations. If the core is a separate region centered on the black hole and accretion disk it would not be surprising if its magnetic field structure were turbulent. In the standing shock model discussed by \citet{mar08}, where the core is assumed to be part of the jet base, to explain the lack of polarization we are asked to believe that this one location in the jet has a turbulent magnetic field structure while the rest of the jet displays a highly organized magnetic field. Although this is possible it would seem unlikely, since this is not a random jet fluctuation but is a feature that is common to all sources. When in a few cases the cores show evidence of polarization, because the nuclear region and the JB are both inside the telescope beam it is impossible to know where it originates. Because most polarization work to date has been done only on a few of the most active sources (BL Lacs) \citep{lis00,gab06,jor07,dar07,dar09} it may not be representative of the majority of radio-loud quasars.

\citet{chi99} have shown that for radio galaxies, both the optical and radio core luminosities are strongly correlated. The tightness found for this correlation indicates that the two spectral bands likely have a common origin. It also seems plausible that not only is the SED reasonably broad, but that at least a strong component of the core radiation, both in radio and optical, must be stable for reasonably long periods of time. However, stability is not something we normally associate with the jet ejection process and this may then be further evidence that a large part of the core radiation is not part of the jet. These investigators also claim that the core radiation is likely to be synchrotron radiation, like that from the jets and lobes, and this may well be correct. However, \citet{jia09} have described a mechanism by which synchrotron radiation could be produced in a core component that is not part of the jet, but instead is centered on the nucleus of the galaxy.

Here we define the jet base as the innermost portion of the jet that lies between the central black hole and the radius set by the half power width of the telescope beam. Before we can understand just what is happening in the central engine of AGNs it is important to know exactly where the radiation from the core originates. It is clear that confusion introduced by the lack of adequate resolution has meant that previous attempts to pinpoint the origin of the AGN core radiation have produced inconclusive results. In what follows we attempt to determine the point of origin of the core radiation from core-dominant, radio loud quasars using an analysis that is unaffected by the present lack of resolution of the region near the central compact object.

\section{Analysis and Results}

VLBA images at 15 GHz of core-dominant, radio-loud quasars reveal the presence of a compact core with an extended, milliarcsec, inner jet structure \citep{kel98}. The jet in many cases is composed of "blobs" separated by gaps. Individual blobs ejected in the jets can be tracked for long periods of time as they move outwards, and this has been done for over 200 blobs \citep{kel04}. In plots of their positions versus time \citep[see their Fig 1]{kel04} there is no evidence that the speed of an individual blob in the inner jet changes significantly as it moves outward.

\begin{figure}
\hspace{-0.5cm}
\vspace{-1.5cm}
\includegraphics[width=8cm]{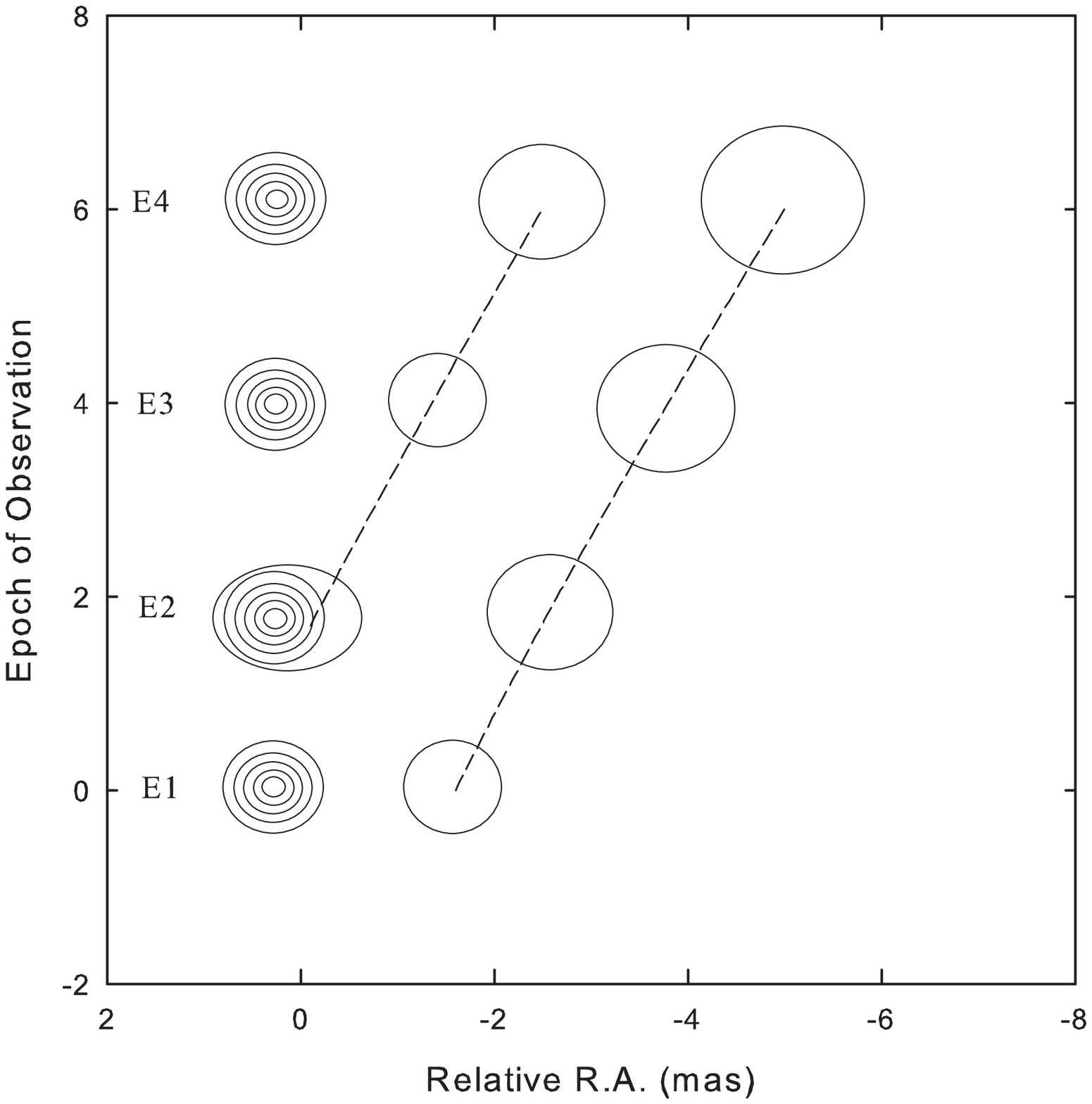}
\caption{{A compact core and inner jet typical of those seen in many radio-loud, core-dominant quasars. There can often be gaps of a few years between successive blob ejections where the flux density in the jet falls below the detection limit. \label{fig1}}}
\end{figure}

 In Fig 1 below, simulated images at 4 different epochs are presented of a source with a strong core and weaker inner jet. At epoch 1 (E1) only the core and one ejected blob are present. At epoch 2 (E2), blob 1 has moved further from the core and the ejection of a new, second blob is imminent. At epoch 3 (E3), the second blob has now moved clear of the core and the first blob has moved still further out. This is, albeit, an over-simplified picture of what has been observed in approximately 130 AGNs monitored by the MAJOVE program \citep{lis09}, but it is adequate for our purpose. In these sources the flux density of the inner jet components is almost always much weaker than that of the core component, with each jet containing on average only $\sim$10 percent of the total flux density \citep{mur93,coo07}. From Fig 1 of \citet{kel04} the average number of blobs per jet is found to be 2.0. Thus the mean core/blob flux density ratio is $\sim20/1$. In a few extreme cases, after a long period in which a series of consecutive ejections have occurred, the flux density from the jet component that is external to the core can become almost as strong as the flux from the core itself.

\subsection{Can the core radiation be explained if it originates entirely in the JB?}

In the MOJAVE monitoring program the angular motions, $\mu$, seen for the ejected blobs is typically $\mu \sim0.5 - 1$ mas yr$^{-1}$ \citep{kel04}, and blob ejections occur typically every 1 to 3 years. Given the beam size of the VLBA (0.3-0.5 milliarcsec) and the observed angular motion, material in the JB can be expected to move away from the unresolved core within a few months, and this has been confirmed by \citet{vil09} who observed delays between optical and  radio of $\sim100$ days in BL Lac. Because of the gaps between ejected blobs, and the above timescales, no more than one new blob is likely to be coincident with the core at any given time. Also, because of the gaps there will be times when there will be no blobs in the JB. If all the core flux was coming from blobs in the JB the core flux density would drop to zero at these times. Since the core is never observed to drop to zero it is clear that a continuous low-level flow must always be present to explain it. Presumably this low-level flow is undetected in the gaps between the blobs further out in the jet because it is below the detection limit.

Since the flux density from the core component is so much stronger than the upper limit set by the flow in the gaps, to explain the core flux this way we also have to assume that the low-level flow must be strongly enhanced while it is in the JB. This enhancement could be due to some mechanism related to stronger magnetic fields in this region, or to the standing shock scenario discussed by \citet{mar08}. \citet{coo07} list the values obtained at 1.4 GHz for the core flux density, S$_{core}$ and the lowest contour level (LC) in mJy beam$^{-1}$, for many of the MOJAVE sources. S$_{core}$/LC values are typically a few hundred to a few thousand. A lower limit to the enhancement factor (EF) required can then be estimated from the ratio of the core flux to the detection limit. From the results of \citet{coo07} we find that an EF $>100$ is required.
 
Thus the core radiation can apparently be explained if it originates in the jet and is mainly due to a continuous, low-level flow that is enhanced by a large factor while passing through the jet base.

\begin{figure}
\hspace{-0.5cm}
\vspace{-1.5cm}
\includegraphics[width=8.0cm]{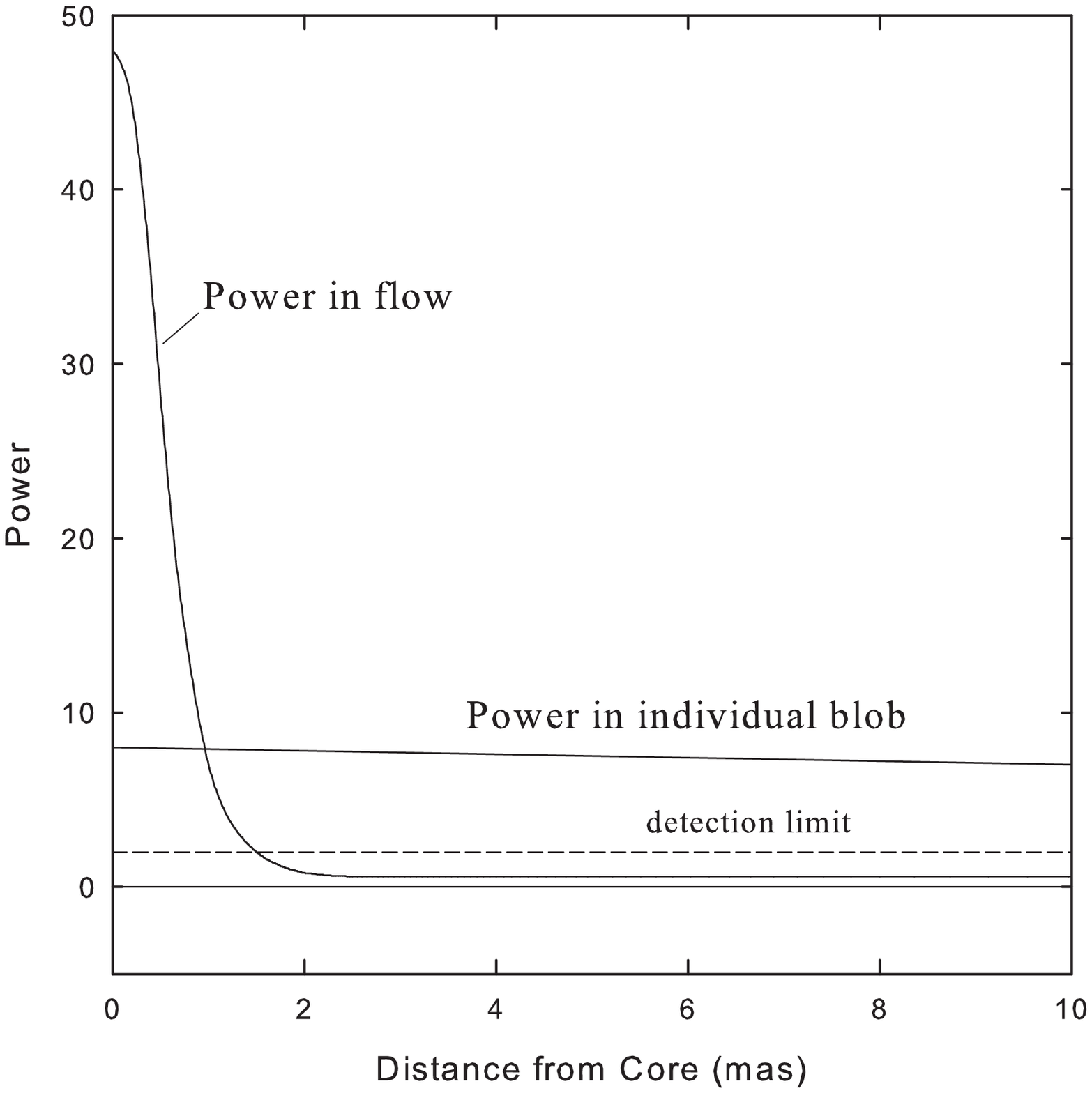} 
\caption{{Plots of radio power in the postulated low-level flow needed to explain the strong core (curved line), and that observed in an individual blob (straight line) versus distance from the central compact object. Because no significant increase is seen in the blob power at radio when it is in the jet base the strong core flux cannot be explained by a low-level continuous flow and it must then originate in a separate component. \label{fig2}}}
\end{figure}

\subsection{Discussion}

Although the above explanation appears to allow the core radiation to originate in the JB, there are problems. Since the ejected blobs are part of the jet flow they must also pass through the same region of high enhancement in the JB as the continuous flow, and they can then be used to determine what the true enhancement factor is. When a new influx of particles is injected into the JB (i.e. a new blob is born) it too should be enhanced by the same factor when it is in this region, assuming that both have reasonably broad energy distributions, as would appear to be the case \citep{chi99}. A lower limit to the amount that the core will be expected to increase when a new blob enters the JB can be estimated from its strength after it clears the core, relative to the lowest contour level. A blob that is a factor of five above the detection limit will then be expected to increase the core flux by at least the same factor while it is in the JB. The core component must then increase by a large factor for at least a period of a few tens of days every time a new influx of particles enters the JB. This large increase should be easily detected in variable-source monitoring programs. In fact, in those MOJAVE movies that include the optical and X-ray fluxes, these bursts are easily seen at both X-ray and optical frequencies. They are strong, last for a few tens of days, and are well correlated with the ejection of new blobs at 14.5 GHz \citep{cha08}. However, at 14.5 GHz the burst amplitudes are much lower. For most sources there is little evidence at 14.5 GHz for an increase in the core flux of more than 10-20 percent with each new influx of particles. This is demonstrated in Fig 2, and it means that \em the enhancement factor at 14.5 GHz in these sources is much too low for more than a small fraction of the core flux to be explained by a continuous, low-level jet flow. \em Furthermore, this result does not suffer from a lack of resolution since it is based on the detection limit set by the gaps between the blobs \em which are well resolved. \em

\begin{figure}
\hspace{-0.5cm}
\vspace{-1.5cm}
\includegraphics[width=8cm]{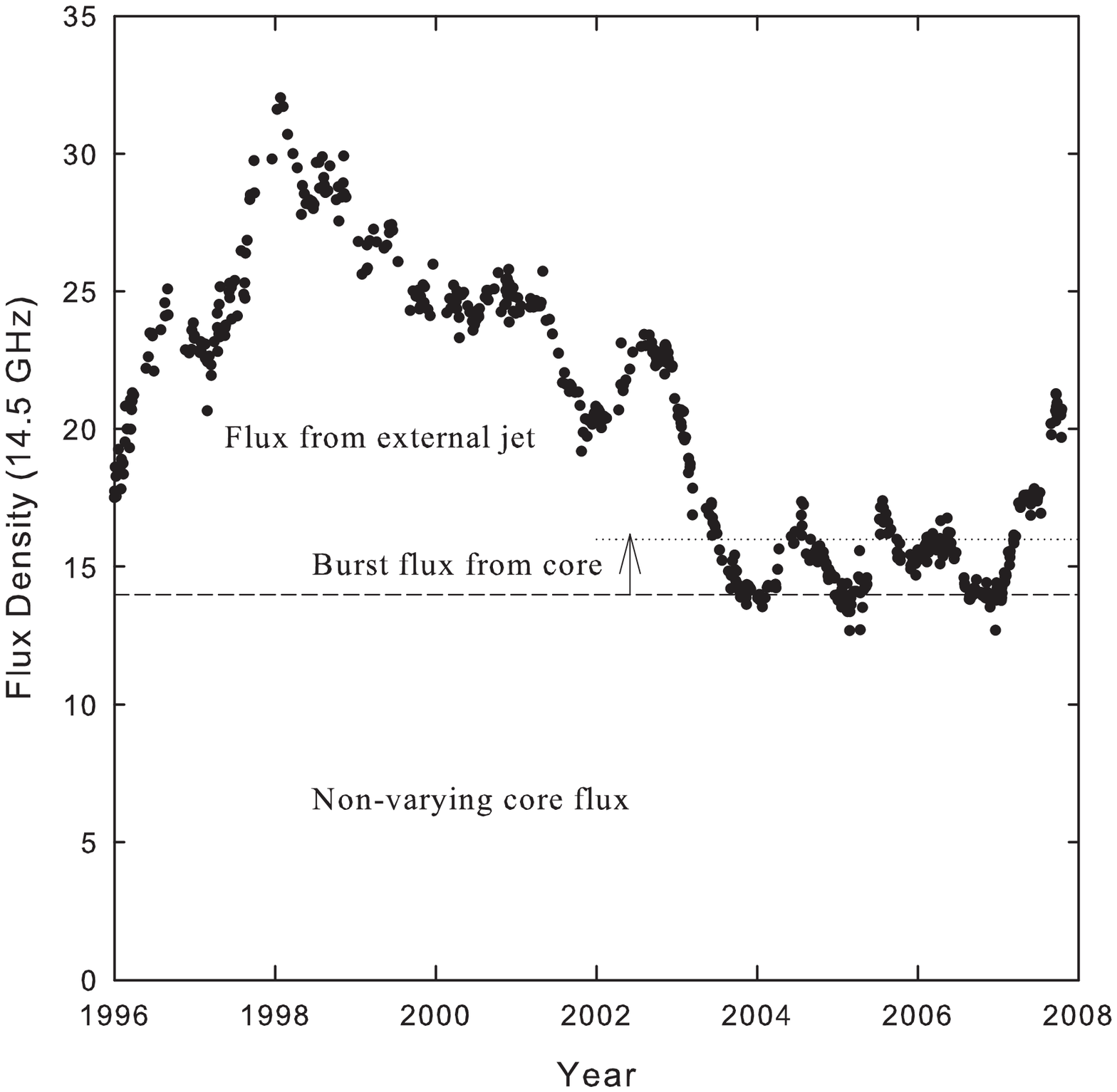}
\caption{{Plot of 14.5 GHz flux vs year for 3C279 from \citet{cha08}. \label{fig3}}}
\end{figure}

To examine this more closely we use the multi-waveband flux monitoring data obtained for 3C279 and discussed by \citet{cha08}. This source is one of only a few very strong radio variables that are known. Fig 11 of that paper shows the X-ray, optical and 14.5 GHz radio flux for this source between y1996 and y2008. First, it is apparent that there are many short-term bursts present at optical and X-ray. As these authors claim, the bursts are well correlated with the ejection of new blobs at 14.5 GHz. Fig 7 of \citet{cha08}, shows how the external jet structure changes with time. The flux from the extended jet outside the core can be seen to increase in y1997, to reach a broad peak near y2001, and then to decrease to a low level by the end of y2003. It then remains low for several years. In Fig 3 here, the 14.5 GHz flux data for 3C279 from \citet{cha08} are plotted versus year. This same variation with time is seen in the total flux in Fig 3. This clearly indicates that the broad peak in the 14.5 GHz flux near y2000 is due to the increase in flux from the \em external \em jet component. The total 14.5 GHz flux of 3C279 then consists of three separate components; 1) a reasonably stable core component near 14 Jy, 2) a low-level, bursting core component associated with the injection of new particles into the JB and, 3) a slowly varying component that fluctuates with the blob ejection activity and originates in the external jet. These are indicated in Fig 3, where the dashed line indicates the stable core component, while everything above it is due to the external jet flux with the short-duration bursts from the JB superimposed. The core flux then consists of two components; 1) a stable component that is unrelated to the jet and, b) a series of short duration burst components associated with the influx of new particles into the jet base. Note that this latter component appears as part of the core component only because the jet base and accretion disk are unresolved. For 3C279, the short duration bursts, indicated by the arrow in Fig 3, represent about 15 percent of the total flux from the core. Thus $\sim85$ percent of the core flux comes from the stable component which must originate in material that is not part of the JB.

This same result can be seen in the BL Lac data \citep{vil09} where the optical bursts are very large ($\sim600$ percent increase in flux) while the increase in the 14.5 GHz flux in each burst is closer to 50 percent with a stable core component near 2 Jy. The observed decrease in burst amplitude with decreasing observing frequency may indicate a decrease in the ambient magnetic field strength with increasing distance from the central compact object, assuming that optical depth plays a role. BL Lac and 3C279 are two of the strongest radio variables known. In less active sources the bursting component, associated with the ejection of new particles into the JB, will therefore be expected to be considerably lower ($<15\%$ of the total core flux), while the stable core flux is $\sim80\%-90\%$ of the total core flux, and, as found above, must be associated with the central accretion area and not the JB.

Unlike most of the previous investigations where the lack of adequate resolution can lead to ambiguous conclusions, in our above determination of the point of origin of the stable component of the core the lack of resolution does not play a role, since it uses the blobs and the gaps between them which are well resolved. Our analysis may also make the standing shock scenario proposed by \citet{mar08} an unlikely candidate for the stable component of the core flux. On the other hand, our conclusions appear to be completely consistent with the model recently discussed by \citet{jia09}.

In summary, the MOJAVE monitoring program and \citet{cha08} have shown that the bursting component of the core flux correlates well with the ejection of new blobs and we can conclude that whether the bursts are at X-ray, optical, or radio, they are therefore associated with a new influx of particles into the jet base. \citet{mur93} and \citet{coo07}, together with the work of \citet{kel04}, have shown that on average the flux from the core is $\sim$10 times stronger than that from the inner jet. We have shown above using the radio data that the blob radiation is not enhanced significantly in the core area and is then unable to explain the strong core flux. It can be concluded from this that the strong, stable, non-flaring component of the core cannot come from the jet base and then must originate in a component associated with the accretion disk, which is likely to be centered closer to the black hole. Although our investigation deals only with the radio frequency radiation, any stable component at X-ray and optical would also be expected in our model to be related to the accretion disk and not the jet base.

 Finally, one of the most obvious characteristics of the radio-loud AGNs discussed here is their flat spectra. When the origin of the core radiation is assumed to be the jet base, the flat core spectra have been explained as arising from a series of self-absorbed synchrotron spectra with shifted peaks \citep{mar77,ban85}. In our model, where the main component of the core radiation is found to come from a stable, quasi-spherical or disk-like region surrounding the accretion disk, the flatness of the spectra still requires explanation. We note in this regard, that flat spectra can be produced in any optically thick environment which is inhomogeneous, such as a spherical or disk-shaped source with a radial gradient in the physical conditions (i.e. magnetic field, relativistic particle density). The argument is the same as for the jets. However, in this case the radiation is expected to be much more stable than if it originated in the jet base. This is then consistent with the stable core component discussed above. Furthermore, since this region must be closer to the central black hole and accretion disk, explaining how the required high-energy electrons survive is easier than if they have to be transported further out into the jet base.

\section{Conclusion}

We have demonstrated here using the core/gap flux ratio and the observed increase in the core flux with each new injection of particles that the enhancement factor in the JB is much too small for more than a small percentage of the core flux to be explained by a continuous jet flow. This means that a large percentage of the total core flux at 14.5 GHz (the stable component) must come from outside the jet base and we conclude that it likely radiates at least quasi-isotropically and is centered close to the position of the black hole. Using the 14.5 GHz flux monitoring data obtained for 3C279 between y1996 and y2008 we find for this source that the stable component of the core radiation that must originate outside the jet base contains $\sim85\%$ of the total core flux. The long-term stability of this component may also indicate that it comes from a region that extends over many Schwarzschild radii. Our result also explains why the position of the core does not move significantly, and why sources that may not presently have an active jet can still have a strong core. Finally, unlike most other attempts to define the point of origin of the AGN core radiation our results are the only ones that are not significantly affected by the fact that the accretion disk and jet base cannot be resolved.

\section{Acknowledgements}

We thank E. R. Seaquist for several valuable comments. 



\begin{thebibliography}

\bibitem[Band and Grindlay(1985)]{ban85} Band, D.L. and Grindlay, J.E. 1985, \apj, 308, 576
\bibitem[Blandford and Znajek(1977)]{bla77} Blandford, R.D., and Znajek, R.L. 1977, \mnras, 179, 433
\bibitem[Blandford and Payne(1982)]{bla82} Blandford, R.D., and Payne, D.G. 1982, \mnras, 199, 883 
\bibitem[Blundell and Kuncic(2007)]{blu07} Blundell, K.M. and Kuncic, Z. 2007, \apj, 668, L103
\bibitem[Chatterjee et al.(2008)]{cha08} Chatterjee, R. et al. 2008, \apj, 689, 79
\bibitem[Cheung et al.(2007)]{che07} Cheung, C.C., Harris, D.E. and Stawarz, L. 2007, \apj, 663, L65
\bibitem[Chiaberge et al.(1999)]{chi99} Chiaberge, M., Capetti, A., and Celotti, A. 1999, \aap, 349, 77
\bibitem[Cooper et al. (2007)]{coo07} Cooper, N.J., Lister, M.L., and Kochanczyk, M.D. 2007, \apjs, 171, 3376 (astro-ph/0701072)
\bibitem[D'Arcangelo et al.(2007)]{dar07} D'Arcangelo et al. 2007, \apj, 659, l107
\bibitem[D'Arcangelo et al.(2009)]{dar09} D'Arcangelo, F.D, et al. 2009, \apj, 697, 985
\bibitem[Falcke et al.(2009)]{fal09} Falcke, H., Markoff, S., and Bower, G.C. 2009, \aap, 496, 77
\bibitem[Gabuzda et al.(2006)]{gab06} Gabuzda, D.C., Rastorgueva, E.A., Smith, P., and O'Sullivan, S. 2006, \mnras, 369, 1596
\bibitem[Jiang et al.(2009)]{jia09} Jiang, Y.-F., Ciotti, L, Ostriker, J.P., and Spitkovsky, A. 2009, preprint astro-ph/09044918
\bibitem[Jorstad et al.(2007)]{jor07} Jorstad, S. et al. 2007, \aj, 134, 799
\bibitem[Jorstad et al.(2009)]{jor09} Jorstad, S. et al. 2009 http://www.bu.edu/blazars/mmpol/ 3c279mov.html
\bibitem[Kellermann et al.(1998)]{kel98} Kellermann, K.I., Vermeulen, R.C., Zensus, J.A., and Cohen, M.H. 1998, \aj, 115, 1295
\bibitem[Kellermann et al.(2004)]{kel04} Kellermann, K.I. et al. 2004, \apj, 609, 539
\bibitem[Krolik(1999)]{kro99} Krolik, H. J. 1999, \em Active Galactic Nuclei: From the Central Black Hole to the Galactic Environment. \em Princeton University Press, Princeton.
\bibitem[Lister and Smith (2000)]{lis00} Lister, M.L., and Smith, P.S. 2000, \apj, 541, 66
\bibitem[Lister et al.(2009)]{lis09} Lister, M.L. et al. 2009, \aj, 137, 3718
\bibitem[Macdonald and Thorne(1982)]{mac82} Macdonald, D., and Thorne, K.S. 1982, \mnras, 198, 345
\bibitem[Marscher(1977)]{mar77} Marscher, A.P. 1977, \apj, 216, 244
\bibitem[Marscher(2008)]{mar08} Marscher, A.P. 2008, in \em Extragalactic Jets: Theory and Observations from Radio to Gamma Ray, \em ASP Conference Series, Vol 386, eds. T.A. Rector and D.S. De Young.
\bibitem[Meier et al.(2001)]{mei01} Meier, D.L., Shinji Koide, and Yutaka Uchida, 2001, Science, 291, 84
\bibitem[Murphy et al.(1993)]{mur93} Murphy, D., Browne, I., and Perley, R. 1993, \mnras, 264, 298
\bibitem[Villata et al.(2009)]{vil09} Villata, M. et al. 2009, preprint in astro-ph/0905.1616

\end{thebibliography}
\end{document}